# Electrostatic Self-assembly :
# A New Route Towards Nanostructures

**J.-F. Berret, P. Hervé, M. Morvan**
Complex Fluids Laboratory, UMR CNRS - Rhodia n°166,
Cranbury Research Center Rhodia 259 Prospect Plains Road, Cranbury NJ 08512 USA

**K. Yokota, M. Destarac**
Rhodia, Centre de Recherches d'Aubervilliers, 52 rue de la Haie Coq,
F-93308 Aubervilliers Cedex France

**J. Oberdisse**
Groupe de Dynamique des Phases Condensées, UMR CNRS – Université de
Montpellier II n° 5581, F-34095 Montpellier France

**I. Grillo and R. Schweins**
Institute Laue-Langevin, BP 156, F-38042 Grenoble cedex 9 FRANCE

Corresponding author: jean-francois.berret@ccr.jussieu.fr
Text written for the ILL Topical brochure on soft matter, 2005

The mechanisms of self-assembly are playing a central role in the field of soft condensed matter. Surfactants are low molecular weight molecules consisting of two antagonist parts, one water-loving (hydrophilic) and one water-repelling (hydrophobic). In aqueous solutions, hundreds of surfactants can stick together in order to minimize the contacts between the hydrophobic parts and the water molecules. As a result, aggregates called micelles are formed (Fig. 1). Depending on the nature of the surfactants, micellar morphologies can be globular, cylindrical or planar [1]. Similar spontaneous mechanisms of association occur for macromolecules which are made from polymer segments showing again antagonist properties with respect to that of the solvent. Amphiphilic diblock copolymers are made of two chains of different nature, one block being hydrophilic, the second being hydrophobic. They are equivalent to surfactants with much higher molecular weights, and the three basic morphologies mentioned previously are recovered with copolymers. Nowadays, block copolymer and their colloidal assemblies are extensively used in applications such as in drug delivery, catalysis and stabilization of colloids and emulsions [3]. With respect to these supramolecular aggregates, it should be noticed that small-angle neutron scattering (SANS) has been the major investigation tool during the last three decades [2].

Surfactant and block copolymers are organic molecules, and their self-assembly in aqueous media is based on the hydrophobic effect. More recently, the controlled association of colloids and macromolecules that would not proceed through the hydrophobic effect has been recognized as a major challenge in the field of physcial-chemistry.

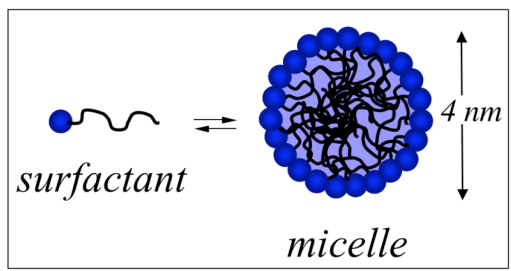

**Figure 1** : *Spherical micelles are supramolecular aggregates made from surfactants.*

The aim is to associate components of different natures, such as organic and mineral or synthetic and biological. Obviously, this type of approach would have a huge impact on applications in material science, nanotechnolgies and in biology.

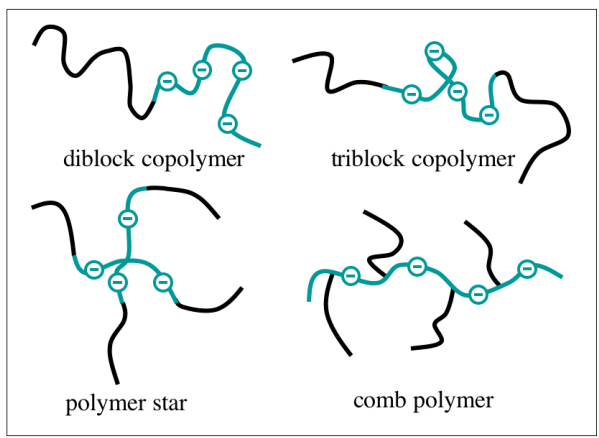

**Figure 2** : *Recent advances in polymer synthesis have allowed to develop new architectures for polyelectrolyte-neutral copolymers. Note that all polymer segments shown here are hydrophilic.*

In the 90's, several international research groups have anticipated that electrostatic interactions could serve as a basis for a novel self-assembly mechanism. Controlled electrostatic self-assembly needs two ingredients to be successful. The first ingredient is a pair of oppositely charged colloids or macromolecules. As a first element of this couple, we use a polymeric chain that carries electrostatic charges along its backbone. These chains, also called polyelectrolytes display a high solubility in water. Polyelectrolytes have strong adsorbing capabilities on surfaces bearing an opposite charge. The second element of the pair is a nanometer-size colloid, which charges are opposite to that of the polyelectrolyte. For instance, Harada and Kataoka used a small and bulky protein, the chicken egg white lysozyme (diameter 5 nm) [4] and Bronich and coworkers a surfactant micelle [5]. When the polyelectrolyte is mixed with oppositely charged colloids, it tends to adsorb on it, due to the attraction between the plus and minus charges. This attraction is strong and usually and Brownian motions are insufficient to desorb the attached species. If the chains are long enough, they can adsorb on several colloids at the same time. A catastrophic process results from this multi-sites binding, leading to a phase separation. Phase separations in solutions of oppositely charged species have been known for almost a century, and serve as basic protocols forwater treatment or in biology for the immobilization of enzymes and purification of proteins. The very existence of a phase separation is indicative that the electrostatic attractions are too strong. For stable and controlled self-assembly, it needs to be counter-balanced by some repulsive forces. This is where modern polymer chemistry enters the game. It provides the second ingredient of the electrostatic self-assembly process, the stabilization.

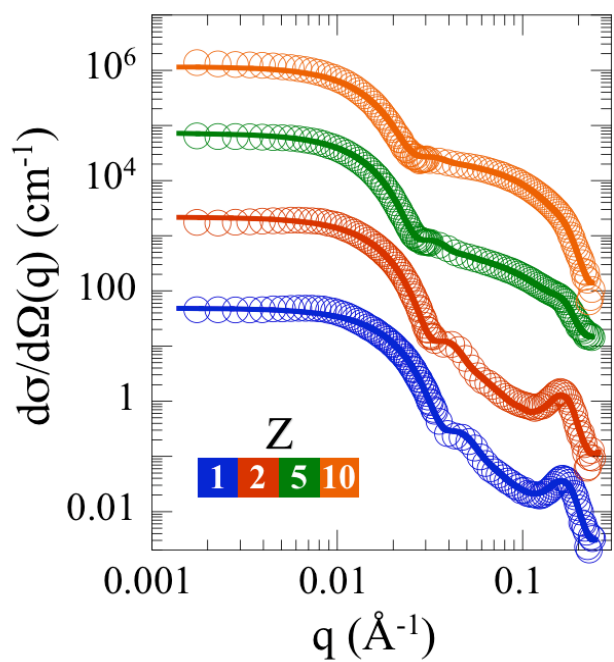

**Figure 3** : *Neutron scattering cross-sections obtained from neutral-polyelectrolyte and surfactant mixed aggregates. The polymers used are the poly(sodium acrylate)-b-poly(acrylamide) diblocks and the surfactants are dodecyltrimethylammonium bromide (DTAB). Poly(sodium acrylate) is negatively charged in water, poly(acrylamide) is neutral and DTAB is positively charged. The molecular weights for the charged and for the neutral blocks are 5000 and 30000 g mol$^{-1}$, respectively. The different scattering curves were obtained at different charge ratios Z between the surfactant and the polymers (Z = 1, 2, 5 and 10). The continuous curves are fits assuming for the microstructure that of Fig. 4.*

Polymer scientists have found recently new synthesis routes to covalently bind a second chain at the extremity of the polyelectrolyte. This second part is neutral, water-soluble and it is aimed to soften the strong attraction mediated by the charges. The authors mentioned previously had used a poly(ethylene oxide) block, the most studied water-soluble neutral polymer, comprising tens or hundreds of monomers. The architecture of such a diblock copolymer is illustrated in Fig. 2a. In more recent studies, the addition of several neutral blocks to the main chain has been made available, yielding a series of novel architectures such as triblocks (Fig. 2b), stars (Fig. 2c) or comb-like (Fig. 2d) polymers.

During the last 3 years, our group has investigated extensively the complexation mechanism between neutral-polyelectrolyte block copolymers, as those depicted in Fig. 2 with oppositely charged species. These species are surfactant micelles, multivalent counterions and inorganic nanoparticles. In the three cases, we have established the thermodynamical phase diagram of these systems, and found broad regions where supramolecular aggregates spontaneously form via electrostatic self-assembly. From earlier works [4-6], it was suspected that these mixed colloids exhibit a core-shell structure. However, their inner structure was unveiled by us only recently, using a combination of light, neutron and x-ray scattering experiments [7-12]. In this respect, the role of the SANS technique was crucial, since it allowed a quantitative determination of the microstructure.

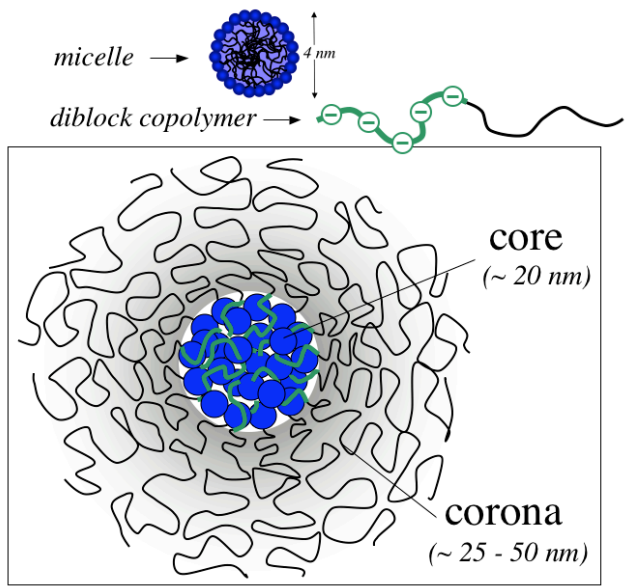

**Figure 4** : *Microstructure assumed for the mixed polymer-surfactant complexes. The mechanism yielding this structure is the electrostatic self-assembly. Such core-shell aggregates were also found using multivalent counterions and inorganic nanoparticles.*

The neutron scattering cross-sections obtained from diblocks and surfactants mixed solutions are shown in Fig. 3, together with fits calculated from the model microstructure of Fig. 4. Here, the copolymer is a poly(sodium acrylate)-*b*-poly(acrylamide) diblock, where the poly(sodium acrylate) is negatively charged in water and poly(acrylamide) is neutral. For the micelle-forming surfactant, we use the C12-cationic dodecyltrimethylammonium bromide. In Fig. 4, the core is described as a complex coacervation micro-phase constituted by densely packed micelles and connected by the polyelectrolyte blocks. The corona is made from the neutral chains and insures the stability of the whole aggregate. Similar types of core-shell colloids were obtained with the multivalent counterions and with the inorganic nanoparticles. In the three cases, the mixed aggregates form spontaneously and have a remarkable long-term stability. We anticipate that the process of electrostatic self-assembly described in this short review could have in the future a large impact in material and biological sciences.

# References

[1] *J.N. Israelachvili*, Intermolecular and Surface Forces, London, Academic Press (1992).
[2] *G. Riess*, Micellization of Block Copolymers, Prog. Polym. Sci. **28**, 1107 - 1170 (2003).
[3] *P. Lindner, T. Zemb*, Neutrons, X-rays and Light : Scattering Methods Applied to Soft Condensed Matter, Amsterdam, Elsevier (2002).
[4] *A. Harada, K. Kataoka*, Novel Polyion Complex Micelles Entrapping Enzyme Molecules in the Core : Preparation of Narrowly-Distributed Micelles from Lyzosome and Poly(ethylene glycol)-Poly(aspartic acid) Block Copolymers in Aqueous Medium, Macromolecules **31**, 288 - 294 (1998).
[5] *T.K. Bronich, A.V. Kabanov, V.A. Kabanov, K. Yui, A. Eisenberg*, Soluble Complexes from Poly(ethylene oxide)-block-Polymethacrylate Anions and N-alkylpyridinium Cations, Macromolecules **30**, 3519 - 3525 (1997).
[6] *M.A. Cohen-Stuart, N.A.M. Besseling, R.G. Fokkink*, Formation of Micelles with Complex Coacervate Cores, Langmuir **14**, 6846 - 6849 (1998).
[7] *P. Hervé, M. Destarac, J.-F. Berret, J. Lal, J. Oberdisse, I. Grillo*, Novel Core-Shell Structures for Colloids Made of Neutral/Polyelectrolyte Diblock Copolymers and Oppositely Charged Surfactants, Europhys. Lett. **58**, 912 - 918 (2002).
[8] *J.-F. Berret, G. Cristobal, P. Hervé, J. Oberdisse, I. Grillo*, Structure of Colloidal Complexes obtained from Neutral/Polyelectrolyte Copolymers and Oppositely Charged Surfactants, Eur. J. Phys. E **9**, 301 - 311 (2002).
[9] *J.-F. Berret, P. Hervé, O. Aguerre-Chariol, J. Oberdisse*, Colloidal Complexes Obtained from Charged Block Copolymers and Surfactants : A comparison between Small-Angle Neutron Scattering, Cryo-TEM and Simulations, J. Phys. Chem. B **107**, 8111 - 8118 (2003).
[10] *J.-F. Berret, J. Oberdisse*, Electrostatic self-assembly in polyelectrolyte-neutral block copolymers and oppositely charged surfactant solutions, Physica B **350**, 204 - 206 (2004).
[11] *J.-F. Berret, B. Vigolo, R. Eng, P. Hervé, I. Grillo, L. Yang*, Electrostatic Self-Assembly of Oppositely Charged Copolymers and Surfactants: A Light, Neutron and X-ray Scattering Study, Macromolecules **37**, 4922 - 4930 (2004).
[12] *K. Yokota, M. Morvan, J.-F. Berret, J. Oberdisse*, Stabilization and Controlled Association of Inorganic Nanoparticles using Block Copolymers, Europhys. Lett., accepted (2004).